# Boosting tunable blue luminescence of halide perovskite nanoplatelets through post-synthetic surface trap repair

*Bernhard J. Bohn,*[1,2,#] *Yu Tong,*[1,2,#] *Moritz Gramlich,*[1,2] *May Ling Lai,*[3] *Markus Döblinger,*[4] *Kun Wang,*[5] *Robert L. Z. Hoye,*[3] *Peter Müller-Buschbaum,*[5] *Samuel D. Stranks,*[3] *Alexander S. Urban,*[1,2,*] *Lakshminarayana Polavarapu,*[1,2,*] *and Jochen Feldmann*[1,2,*]

[1] Chair for Photonics and Optoelectronics, Department of Physics and Center for NanoScience (CeNS), Ludwig-Maximilians-Universität München, Amalienstr. 54, 80799 Munich, Germany
[2] Nanosystems Initiative Munich (NIM), Schellingstr. 4, 80799 Munich, Germany
[3] Cavendish Laboratory, JJ Thomson Avenue, Cambridge CB3 0HE, United Kingdom
[4] Department of Chemistry, Ludwig-Maximilians-Universität München, Butenandtstr. 5-13 (E), 81377 Munich, Germany
[5] Lehrstuhl für Funktionelle Materialien, Physik Department, Technische Universität München. James-Franck-Str. 1, 85748 Garching, Germany
[#] equal contribution
[*] corresponding authors: urban@lmu.de, l.polavarapu@lmu.de, feldmann@lmu.de

ABSTRACT:

The easily tunable emission of halide perovskite nanocrystals throughout the visible spectrum makes them an extremely promising material for light-emitting applications. Whereas high quantum yields and long-term colloidal stability have already been achieved for nanocrystals emitting in the red and green spectral range, the blue region currently lags behind, with low quantum yields, broad emission profiles and insufficient colloidal stability. In this work, we present a facile synthetic approach for obtaining two-dimensional $CsPbBr_3$ nanoplatelets with monolayer-precise control over their thickness, resulting in sharp photoluminescence and electroluminescence peaks with a tunable emission wavelength between 432 and 497 nm due to quantum confinement. Subsequent addition of a $PbBr_2$-ligand solution repairs surface defects likely stemming from bromide and lead vacancies in a sub-ensemble of weakly emissive nanoplatelets. The overall photoluminescence quantum yield of the blue-emissive colloidal dispersions is consequently enhanced up to a value of 73±2 %. Transient optical spectroscopy measurements focusing on the excitonic resonances further confirm the proposed repair process. Additionally, the high stability of these nanoplatelets in films and to prolonged UV light exposure is shown.

Halide perovskites in the form of bulk thin films have attracted significant interest owing to rapid rises in photovoltaic efficiencies.[1-5] Nanocrystals (NCs) of the same material have been garnering strong interest since 2014 for light emission applications due to high photoluminescence quantum yields (PLQYs) approaching unity and an emission wavelength that is tunable throughout the visible range by changing the halide ion.[6-14] Two of the important limitations currently hindering the widespread use of this material, are i) a very low efficiency of chloride-containing perovskite NCs, and thus only poor performance in the blue spectral range and ii) a tendency of mixed-halide perovskite to phase segregate into domains with different halide ion contents and resulting bandgaps, consequently shifting the emission wavelength strongly during operation of LEDs.[15-20] While most of these NCs are either bulk-like or only weakly confined, several studies have shown strong quantum confinement in perovskite NCs as a further viable method to tune the emission wavelength especially towards the blue-spectral range.[21-23] In particular, two-dimensional (2D) nanoplatelets (NPls) have been demonstrated with an atomic-precision control over their thickness down to a single monolayer (ML) and large exciton binding energies, an appealing characteristic for light-emitting applications.[15, 24-30] These NPls are unfortunately strongly susceptible to surface defects due to a large surface-to-volume ratio, rendering their QYs typically quite low.[18, 23, 25, 26, 31] It was often observed that the PLQY of both hybrid and inorganic perovskite NCs drastically decreases with the change in morphology from cubic to platelet form.[23, 24, 26, 32] For instance, the PLQY of $CsPbBr_3$ NCs drops from 78 % to 31 % with a change of morphology from cubic to thick NPls.[23, 24] Additionally, it is difficult to produce NPls of only a single thickness using current synthesis routes, and the resulting NPls tend to have problems with long-term stability.[18, 26, 33-35]

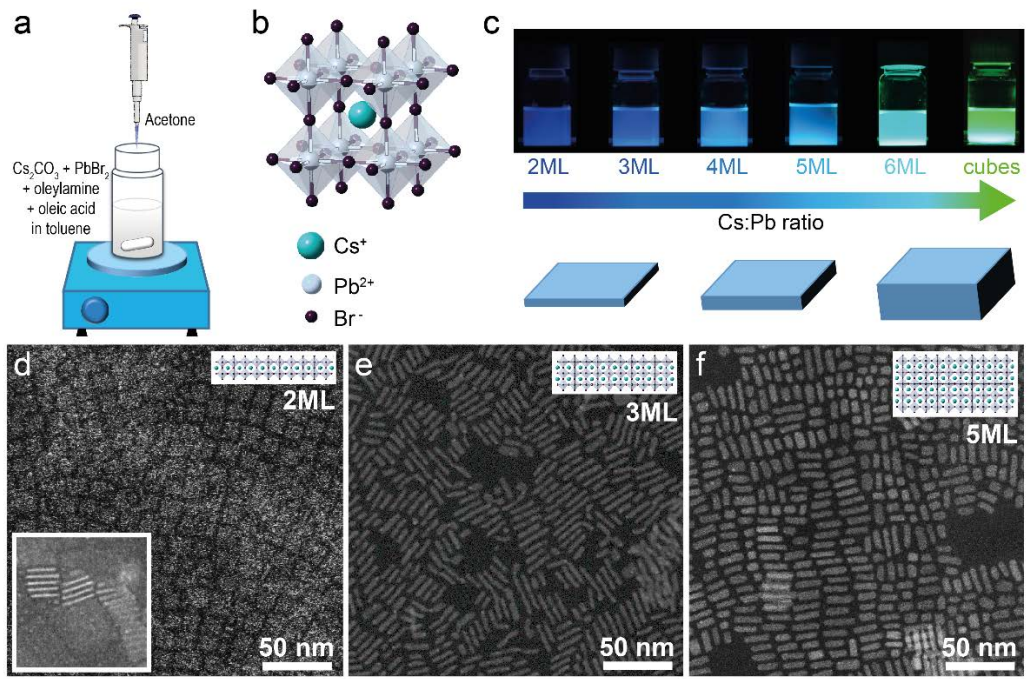

Figure 1. (a) Schematic illustration showing the synthesis of $CsPbBr_3$ nanoplatelets by reprecipitation of precursors using acetone. (b) Cubic crystal structure of $CsPbBr_3$ perovskite NCs. (c) Photographs of vials containing the colloidal NPl dispersions (obtained with increasing Cs:Pb ratio) illuminated with UV-light. A steady redshift with increasing ML thickness can be observed. (d)-(f) HAADF-STEM images of 2ML, 3ML and 5ML NPls.

In this work, we present a simple room temperature synthesis method for obtaining blue-emitting colloidal $CsPbBr_3$ NPls with precisely controllable thickness from 2 to 6 MLs. All NPls exhibit clear excitonic peaks in absorption spectra – a result of the increased exciton binding energies in these confined nanostructures – as well as narrow photoluminescence (PL) and electroluminescence (EL) emission. We show that the PLQYs of these NPls can be significantly enhanced through a post-synthetic treatment by the addition of a $PbBr_2$-ligand solution. The PLQYs of these NPls reach up to 73±2 % with the emission color retained, making this NPl series one of the most efficient perovskite blue light-emitters reported to date.[36] Transient absorption spectroscopy (TAS) and time-resolved (TR)-PL studies reveal that the enhancement affects a sub-ensemble of the NPls for which surface defects – initially hindering radiative recombination – are repaired, enhancing the overall emission of the NPl dispersions. Additionally, we show the enhanced stability of the NPls even under prolonged UV exposure. The synthesis of $CsPbBr_3$ NPls applies the concept of reprecipitation[37] with thickness control achieved by regulating the molar ratio of the precursors, as in our previous studies (see Figure 1a-c).[11, 24] Briefly, a Cs-oleate precursor with a specific Cs:Pb ratio for each NPl thickness (cf. Supporting Information Table S1) was added into a toluene solution containing a $PbBr_2$-oleylamine/oleic acid precursor at room temperature under vigorous stirring (Figure 1a). After 5s, a large amount of acetone was added to initiate formation of the NPls. The reaction was then run for 1 min under continuous stirring, after which the mixture was centrifuged for 3 min and then redispersed in 2 ml of hexane. The thickness of the NPls was controlled by varying the ratio of the Cs:Pb precursors and the amount of acetone added. The details of the synthesis are given in the Supporting Information (Materials and Methods). The resulting dispersions show a strong difference in color, especially under UV-light, with a clear shift from blue to green with increasing Cs:Pb ratio (see Figure 1c). This is indicative of quantum confinement effects and very similar to our previous findings on hybrid organic-inorganic perovskites, where the MA:Pb ratio was controlled.[24] In these two approaches the amount of the A-site cation (here: $Cs^+$, previously $MA^+$) is decreased with respect to the other precursors, limiting the amount of perovskite that can be formed and the resulting nanocrystals shrink in size or more specifically in thickness. Thus, NPls with thicknesses down to a single monolayer can be formed. However, one drawback of our previous approach was that the syntheses tended to form NPls dispersions of mixed thicknesses leading to broad PL emission. To investigate the monodispersity of the NCs in our new findings, we employed annular dark field scanning transmission electron microscopy (ADF-STEM) as shown in Figure 1d-f. For the lowest Cs:Pb ratio, we find regularly sized, low contrast square-shaped NCs with side lengths of 14±4 nm (Figure 1d). The low-contrast signifies very thin NPls, which is confirmed by images of NPls standing on their side (see inset), from which we measured a thickness of 1.2±0.1 nm. One ML in this case signifies a 2D arrangement of corner-sharing $[PbBr_6]^{4-}$ octahedra with a thickness of 0.59 nm (Figure 1b).[11, 12] Therefore, the measured thickness likely corresponds to NPls comprising two MLs. As the Cs:Pb ratio increases, the thickness of the NPls also increases (2.0±0.3 nm and 2.9±0.3 for the 3ML and 5ML NPls, respectively. See Figure 1d-f and also Figure S1 for a detailed analysis of the thickness). Additionally, we find that the lateral size is nearly constant for all of the NPl samples (cf. Figures S1, S2). ADF-STEM imaging and corresponding fast Fourier transform (FFT) patterns confirm the high level of crystallinity of the NPls, which exhibit an orthorhombic or cubic crystal structure (cf. Figure S5). Additionally, inferred from the higher optical contrast, we observe a strong tendency of the NPls to arrange in stacks along the substrates, signifying a high uniformity in the lateral size of the nanoplatelets.[38] This stacking does not happen in the dispersions, as derived from the lack of any precipitate forming and nearly no scattering observed in the UV-VIS spectra. From these analyses, it seems as if the dispersions contain NPls each with only one specific thickness ranging from 2 - 6 MLs. This is further confirmed by X-ray diffraction (XRD) analysis. Additionally to the bulk $CsPbBr_3$ 00l peaks, which signify a high degree of orientation of the NPls, peaks resulting from stacking of the NPls are observed (cf. Figure S7). Using these, we obtain spacing values between individual NPls, which correspond well to the expected thickness of the inorganic layer including a ligand length of 1.5nm surrounding the NPls.

In order to confirm this, we turned to optical spectroscopy, probing the linear PL and absorption spectra of each NPl dispersion (Figure 2a,b). Each PL spectrum clearly comprises only a single narrow peak (full width at half-maximum (FWHM) down to 11 nm), with the central peak position blueshifting from 515 nm to 432 nm with decreasing Cs:Pb ratio in the precursor (Figure 2a). The Cs:Pb ratio used to obtain the 515 nm emission wavelength is comparable to those we used previously to obtain weakly quantum-confined nanocubes.[11, 39] The absorption spectra of all NPl samples show a clear absorption onset, with small Stokes shifts of 6 - 10 nm (Figure 2a). Additionally, as the Cs:Pb ratio is reduced, a clear excitonic absorption peak can be observed and for the lowest ratios, additionally a step-like increase resembling the continuum onset of a 2D semiconductor (Figure 2b). With the spectra being acquired at room temperature, the emergence of the excitonic resonance signifies that the exciton binding energy $E_B$ increases with decreasing Cs:Pb ratio, to an extent that the thermal energy at room temperature is progressively less able to dissociate excitons. As it can be seen in Figure 2b additional peaks emerge at higher photon energies, however, these stem from transitions between higher energetic bands and are not considered further.[40] Similar widths of excitonic absorption peaks and PL emission spectra, small Stokes shifts and the lack of additional peaks in the PL spectra all support the hypothesis that the dispersions actually comprise NPls of only a single thickness. Accordingly, we confirm the correlation of the observed PL peaks and corresponding dispersions to NPls with a ML thickness of n = 2 up to n = 6.[12]

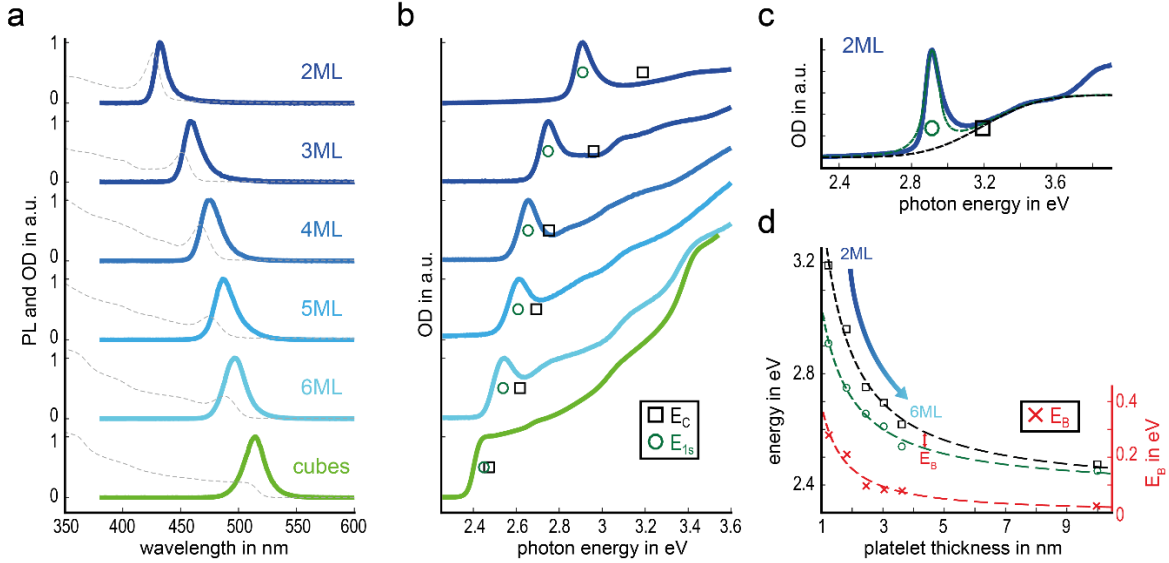

**Figure 2.** (a) Photoluminescence (solid lines) and absorption (dashed lines) spectra of NPl colloids for varying NPl thickness. Spectra of weakly-confined nanocubes are also shown for comparison. (b) Absorption spectra normalized to the excitonic peak. Green circles indicate the energetic position of the exciton transition $E_{1s}$ and black squares show the corresponding continuum absorption onset $E_C$ for each sample. (c) Model (green dashed line) applied to extract these positions from the experimental data (solid curve, in this case 2ML) using the excitonic peak and the absorption onset of the continuum (black dashed line). (d) The obtained values for $E_C$ and $E_{1s}$ versus the NPl thickness. Their difference is the exciton binding energy $E_B$ (red crosses), which increases strongly with decreasing NPl thickness. Dashed lines are added as guides to the eye.

The observed blueshift in the PL spectrum with decreasing Cs:Pb ratio stems from quantum-confinement in the NPls. This effect generally results in the increase of the continuum absorption onset $E_C$ given by the respective band gap energy of the bulk crystal and the confinement energies of the electron and hole, $E_C = E_G + E_e + E_h$. The exciton binding energy $E_B$ increases simultaneously. As these effects actually shift the resulting PL emission in opposite directions, the shift of $E_C$ must be larger than the increase of $E_B$ in order to induce the observed blueshift. To confirm this, we have modelled the absorption spectra using Elliot's model (Figure 2c). Using a method similar to that suggested by Naeem et al. for 2D platelets,[41] we can assign the continuum absorption onset energy $E_C$ (black squares) and also the energetic position of the dominant 1s excitonic transition $E_{1s}$ (green circles) to each of the samples in Figure 2b by modeling the contribution of the excitonic peak and the lowest band of the continuum transitions. These values are important, as their difference is exactly the exciton binding energy: $E_B = E_C - E_{1s}$. This model can be applied through a simple modification to 2D and 3D semiconductors, which apply for the 2 ML and the cube case, respectively. For all other NPls, the actual structure is somewhere in between, as they are only "quasi-2D" due to their finite thicknesses. Yet, they are also not three-dimensional due to a significant degree of quantum confinement in one direction. We were able to extract values for all NPl thicknesses, which we have plotted in Figure 2d. Clearly, both energies decrease with increasing NPl thickness. The exciton binding energy $E_B$, being the difference between the two curves can be seen to increase as the NPl thickness is reduced, reaching a value of 280 meV for the 2ML case. This is nearly 10 times the value for the cubes of only 30 meV, which corresponds well to previously described values.[42-45] This strong change can be understood when comparing the excitonic Bohr radius of bulk $CsPbBr_3$, reported to be approximately $a_B$ = 7 nm, with the NPl thickness. Clearly, the 2 to 6 ML NPls are far thinner than this, thus they are in the limit of strong quantum-confinement, while the cubes with a side length of about 10 - 12 nm are only in the weak limit.[22] Previously it has been shown that the extraction of $E_B$ by analysis of absorption data is prone to fitting ambiguities in the case of bulk perovskite crystals[46] and other approaches have been applied to extract this value.[42, 47] However, here in the case of strong confinement for the NPl dispersions with clearly separated excitonic and continuum absorption, the fitting of absorption data is significantly more reliable.

While the synthesis yields perovskite NPls with uniform thickness and narrow PL emission spectra, the intensity, especially for the thinnest NPls tends to be quite low. In fact, for the 2ML and 3ML samples we determine their PLQY to be only around 7±1 % and 9±1 %, respectively. Such low values have commonly been observed and are likely a result of the existence of surface trap states and the high surface-to-volume ratio of these 2D nanostructures.[23, 25] To address this issue, we have developed a post-synthetic treatment of the colloidal NPls obtained from the previously described synthesis in order to passivate the trap states on the NPl surface. We discovered that by adding a $PbBr_2$-ligand (oleic acid and oleylamine) hexane solution to the NPl dispersion, we can dramatically enhance their PLQYs. Immediately after addition, the PL of the samples is strongly increased, so that the dispersions appear deeply blue even under ambient lighting (see Figure S9). It turns out that the PL intensity of each NPl dispersion can be significantly enhanced, as shown in Figure 3a. The PL peak maxima and absorption features were nearly constant, with some samples showing only minor blueshifts (10-15 meV) after the enhancement. This could be due to an increasing of the dielectric constant of the enhanced samples and a blueshift of the excitonic resonance position due to the increased Coulomb screening. These results indicate that the treatment enhances the PL efficiency of the colloidal NPls without affecting their thickness or uniformity, as confirmed through ADF-STEM imaging and XRD measurements (cf. Figures S3 and S8). Furthermore, electron diffraction measurements show that the crystallinity of the NPls is significantly increased in the enhanced samples (see Figures S4 and S5). Energy-dispersive X-ray (EDX) spectroscopy shows additionally that the ratio of Br:Pb and Pb:Cs ions in the NPls increases as a result of the enhancement (see Supporting Information for details, Figure S6). To investigate the mechanism, for comparison, we added a similar solution without the $PbBr_2$ (only the organic ligands in hexane) to the NPl dispersions. As in our previous work, we find that the dilution and additional ligands tend to decompose the NPls, in this case mainly into their original precursors, as determined through PL spectroscopy (see Figure S10). Furthermore, we applied various additional bromide-salt-ligand solutions ($SnBr_2$, KBr, NaBr) in the enhancement step. All of these lead to an enhancement of the PL, albeit of smaller magnitude than for the $PbBr_2$-solution (see Figure S11). These results together imply that Br and Pb vacancies at the surface of the nanocrystals are filled by the enhancement solution, with the ligands serving to passivate any uncoordinated surface atoms (see Figure 3b).

The enhancement is effective for all thicknesses from 2 to 6 ML with PLQY values of 7±1 % and 42±3 % for the 2 and 6 ML samples before and 49±3 % and 73±2 % after the enhancement process (Figure 3c). The PLQY values of 49±3 % (2ML) and 60±4 % (3ML) are one of the highest reported PLQYs of any perovskite material for an emission wavelength shorter than 470 nm. Interestingly, we observe that the degree of enhancement

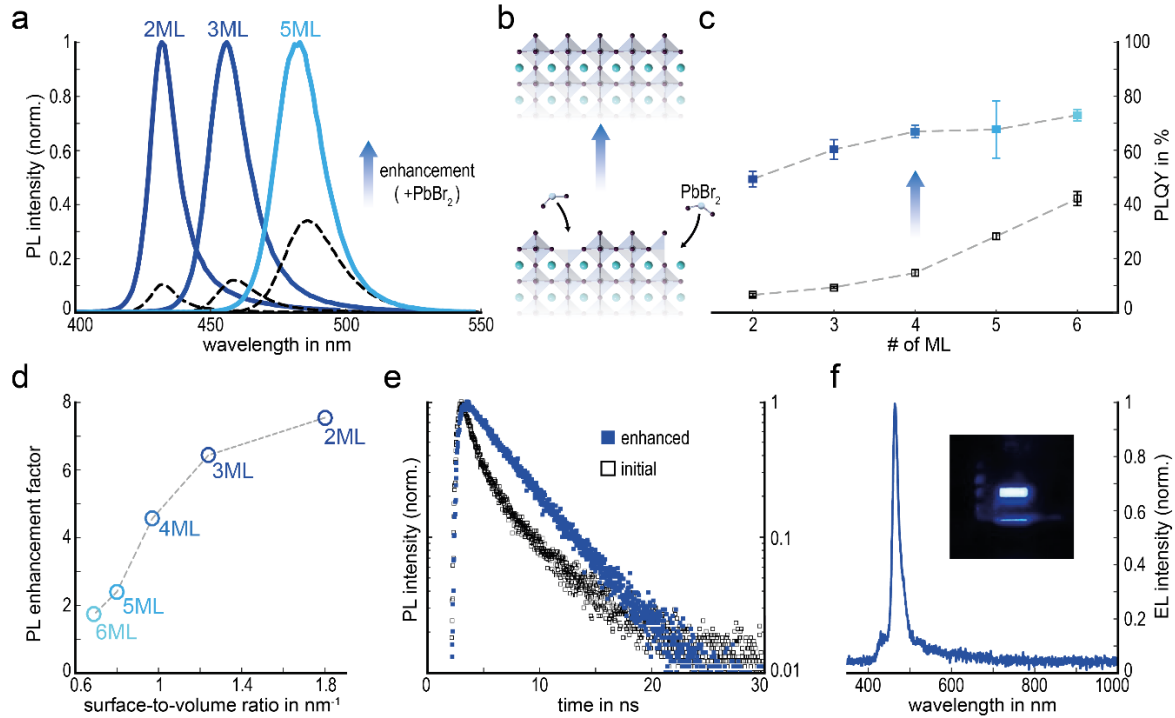

**Figure 3.** (a) PL spectra of initial NPl colloidal dispersions (black dashed lines) and enhanced dispersions (normalized, solid lines). (b) Scheme for the repair process of surface defects initiated by a chemical post-treatment with a PbBr$_2$ solution. (c) PLQYs of the NPL dispersions pre-enhancement (open black squares) and post-enhancement (full blue squares). n ≥ 6 values have been averaged for each data point. Dotted lines are guides to the eye. (d) PL enhancement factor as a function of the surface-to-volume ratio. (e) TR-PL of a 3ML NPl dispersion before (open black squares) and after enhancement (full blue squares). (f) Electroluminescence (EL) spectrum of a light-emitting diode (LED) based on the enhanced 3ML NPls. Inset: photo of the working LED.

is strongly dependent on the thickness of the sample, with the PLQYs improving by a much larger factor for the thinnest platelets (Figure 3d). In fact, there seems to be a strong monotonically-increasing dependence of the enhancement factor on the surface-to-volume ratio. The enhanced NPls also turn out to be extremely stable, as colloidal dispersion retain 90 % of their PL intensity after 10 days (cf. Figure S12) and films obtained through spin-coating retained more than 85 % of their PL intensity even after continuous illumination with UV-light for six hours (see Figure S13). Additionally, when investigating the PL lifetimes of the NPl samples, we find that not only does the lifetime increase after the enhancement procedure (*e.g.* $\tau_{1/e}$ from 1.7 to 4.1 ns for the 3ML sample), but the shape of the decay also strongly changes (see Figure 3e, cf. Supporting Information including Figures S14 and S15). While the decay curve of the initial, non-enhanced samples can only be reproduced through a multiexponential function, the decay of all the enhanced NPl samples becomes monoexpontential. Finally, we demonstrate that the passivation route can yield blue EL when incorporated in a light emitting-diode (LED) device structure (Figure 3f). The LED device based on the 3ML NPls shows pure blue emission with a single, narrow EL peak (FWHM < 20 nm) centered at 464 nm, close to that of the PL (see Supporting Information for more details on the device fabrication and performance, Figure S16). The combination of a narrow emission linewidth and a pure blue color demonstrate their strong potential for use in high-performance LED applications.[19, 48, 49]

To get a deeper understanding of this enhancement procedure one has to consider that semiconductor nanocrystals are known to be highly prone to surface traps, as the atoms/ions sitting at the edges are undercoordinated, leaving dangling bonds.[50] For larger perovskite nanocrystals in the unconfined or weakly quantum confined regimes, surface passivation is highly effective, often leading to extremely high PLQY values approaching unity. On the other hand, strongly confined NCs, for example 2D NPls, tend to have significantly lower QYs. One reason for this could be due to them possessing a higher surface-to-volume ratio, which could render them more susceptible to surface defects. Additionally, the reduced thickness also leads to a reduced dielectric screening, as witnessed by the strongly increased exciton binding energies. This will likely also increase scattering rates of excitons with defects and polar optical phonons, enhancing nonradiative recombination.[28] Such is the case for the NPls obtained through the synthesis before the enhancement step. As a side note, the enhancement process also works for the nanocubes, even though the enhancement factor is far less due to the extremely high QYs of the initial nanocubes (approx. 92 %). This supports the notion of surface traps being the reason for the reduced QYs for thinner NPls. Interestingly, the shape of the PL decay suggests that the NPls are split into two sub-ensembles (see Figure 4a). One ensemble without surface defects and a high QY and one ensemble with surface defects and strongly reduced or no emission. The enhancement renders a larger fraction of the NPls emissive, and thus the PL decay becomes dominated by excitonic recombination and not by non-radiative, trap-mediated decay. This also explains why the enhancement is more effective for the thinner NPls, as surface defects affect them more, increasing a chance that a NPl will be non-emissive.

In order to validate this model and the dramatic increase in the PLQY, we employ transient absorption spectroscopy (TAS). The NPls in solution are spaced relatively far apart and so do not exchange excited charge carriers. Thus, in solution, each NPl can be considered independent of the other NPls. If a NPl in the ensemble is photoexcited, it can contain one or multiple electron hole pairs, which rapidly relax down to the 1s exciton state. For the case of multiple excitations, fast non-radiative exciton-exciton annihilation dominates the recombination of the charge carriers,[51, 52] until only one exciton per photoexcited platelet remains. Now, in a defect-free NPl the excited electron hole pair tends to recombine radiatively (light colored platelet in Figure 4a), whereas the excitation in a NPl with defects has the possibility to recombine radiatively by sending out a photon or non-radiatively via the defect-induced trap states (dark platelet in Figure 4a).

Figure 4b,c show the steady state absorption of the 3ML sample and the respective TAS spectra for an excitation at 400 nm (≈ 3.1 eV) with a dominant bleaching peak at around 452 nm, corresponding to the excitonic peak in the steady state absorption, well below the continuum onset $E_C$. To study the effect of the post-synthetic enhancement on the dynamics of the differential absorption at the energetic position of the exciton, we look at three different colloidal dispersions of 3ML NPls: one before enhancement (initial), one with full enhancement (enhanced), and one where only half the necessary amount of $PbBr_2$ was added and a reduced enhancement is observed (partially enhanced). Figure 4d,e show the transients of differential transmission at the energetic position of the exciton for the initial and the enhanced case.

As proposed above and shown in Figure 4a, we consider there to be two sub-ensembles of platelets in each NPl dispersion. As a consequence the transients in Figure 4d,e represent a superposition of the transient absorption signal of the two sub-ensembles for both cases, the initial and the enhanced sample. As indicated by the black lines, these two transients behave very similar on the very short ($t_{delay}$ < 50 ps) and very long time scales ($t_{delay}$ > 500 ps), however, a significantly different behavior can be observed in the intermediate time regime. With the long timescale process likely being due to recombination, as it is describable through a single exponential function (excitonic recombination), we subtract it from the curves in order to compare the two shorter timescale processes.

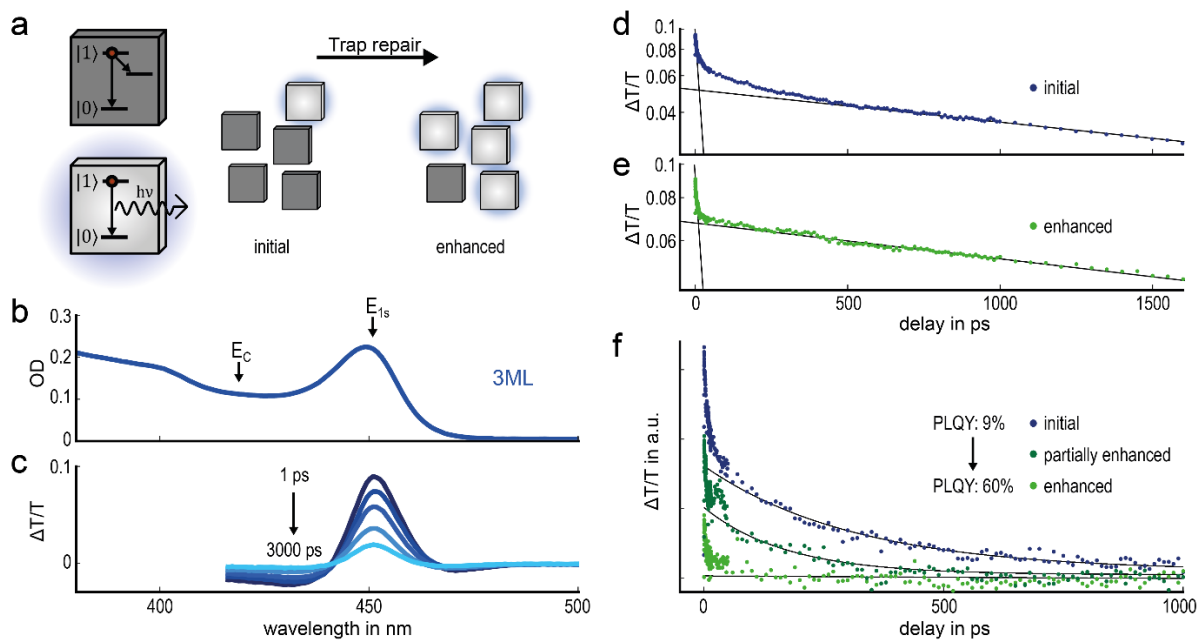

**Figure 4. (a)** Scheme depicting the post-synthetic trap repair and concomitant increase of the number of defect-free nanoplatelets that exhibit predominantly radiative recombination. **(b)** Steady-state absorption spectrum and **(c)** differential absorption spectra at delay times ranging from 1 to 3000 ps after excitation for the 3ML sample. **(d)** Transients of differential absorption at the energetic position of the exciton, as shown in (b) for an ensemble of (d) an initial and of **(e)** an enhanced 3ML sample. **(f)** Transients after subtracting the late time decay corresponding to excitonic recombination. The initial samples show exciton trapping in non-emissive NPls with surface defects as an additional process, which is slower than exciton-exciton annihilation, but faster than excitonic recombination.

After subtracting the late time component, the bleaching decay curves show two main processes happening in the first nanosecond after excitation (see Figure 4f). An initial very fast decay is observed, which we attribute to non-radiative exciton-exciton annihilation, as it exhibits a strong dependency on the excitation density. During the time of the second decay component (50 ps < $t_{delay}$ < 500 ps) a large part of the excitation is lost in the initial sample (dark blue). This amount is substantially reduced in the partially enhanced sample and not observable in the fully enhanced sample. Therefore, we conclude that this decay component stems from a decay channel not present in the enhanced NPls – namely the non-radiative recombination via surface trap states in the sub-ensemble containing defects. This is strong support for the concept of the reduction of surface traps by post-synthetic enhancement with $PbBr_2$ leading to a larger sub-ensemble of defect-free NPls exhibiting strong radiative recombination.

In conclusion, we have shown a facile synthetic approach to obtain colloidal $CsPbBr_3$ NPls with an atomic layer level control of the thickness and with high PLQYs. The NPls, which range in thickness from 2 - 6 MLs all exhibit narrow PL emission in the blue color range as well as sharp excitonic absorption peaks. Through linear optical spectroscopy we confirm that all NPls lie in the strong quantum confinement regime with exciton binding energies up to 280 meV. The high PLQYs of the NPls are achieved through a post-synthetic treatment of the NPl dispersions with a $PbBr_2$-ligand addition. TR-PL and TAS substantiate our proposition that this strong increase in PL is induced by surface trap repair of bromide and lead vacancies in the NPls. This is also the reason why the enhancement process affects the thinnest platelets with a large surface-to-volume ratio in the strongest fashion. This renders the post-processed colloidal NPls exhibiting a high PLQY in the range of 50 % - 75 %, amongst the most efficient, tunable blue-emitting perovskites to date. As shown in a proof-of-concept experiment these NPls could be ideal candidates for blue emitting LEDs, opening the door for all-perovskite white LEDs and other light-emitting applications.

## SUPPORTING INFORMATION:

Materials and Methods for Synthesis and the Enhancement procedure; Morphological and structural characterization: Transmission electron microscopy (TEM), Electron diffraction, High-resolution annual darkfield scanning tunneling electron microscopy (HR-ADF-STEM), Energy-dispersive X-ray (EDX) spectroscopy, X-ray Diffraction (XRD); Optical Characterization: UV-Vis, PL spectroscopy, time resolved (TR) PL spectroscopy; LED fabrication and electroluminescence (EL) measurements.

## ACKNOWLEDGEMENTS:

This research work was supported by the Bavarian State Ministry of Science, Research, and Arts through the grant "Solar Technologies go Hybrid (SolTech)," by the European Research Council Horizon 2020 ERC Grant Agreement No. 759744 - PINNACLE (A.S.U.), by the China Scholarship Council (Y.T. and K.W.), by the European Union's Horizon 2020 research and innovation program under the Marie Skłodowska-Curie Grant Agreement COMPASS No. 691185 and by LMU Munich's Institutional Strategy LMU excellent within the framework of the German Excellence Initiative (A.S.U. and L.P.). R.L.Z.H. would like to acknowledge support from Magdalene College, Cambridge. S.D.S acknowledges funding from the European Research Council (ERC) under the European Union's Horizon 2020 research and innovation programme (grant agreement number 756962), and the Royal Society and Tata Group (UF150033). B.J.B. and Y.T. contributed equally to this work. The authors would like to thank Sebastian Rieger for helpful discussions.

TOC Graphic:

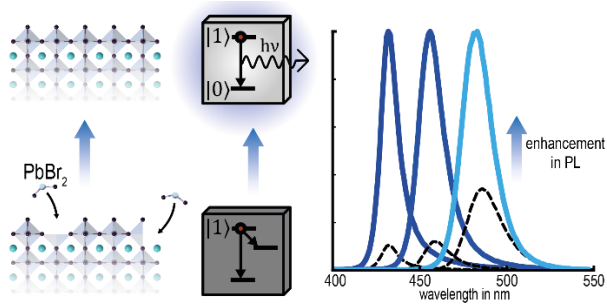